\documentclass[twocolumn,showpacs,preprintnumbers,amsmath,amssymb]{revtex4}
\usepackage{epsfig}
\usepackage{graphicx}
\usepackage{dcolumn}
\usepackage{bm}
\include{epsfig}
\usepackage[colorlinks=true,linkcolor=blue]{hyperref}%

\newcommand{\be}{\begin{equation}}
\newcommand{\ee}{\end{equation}}

\newcommand{\bea}{\begin{eqnarray}}
\newcommand{\eea}{\end{eqnarray}}

\newcommand{\al}{\alpha}
\newcommand{\bt}{\beta}
\newcommand{\gm}{\gamma}
\newcommand{\Gm}{\Gamma}
\newcommand{\dl}{\delta}

\newcommand{\et}{\eta}

\newcommand{\lm}{\lambda}

\newcommand{\rh}{\rho}

\newcommand{\sg}{\sigma}

\newcommand{\ch}{\chi}

\newcommand{\om}{\omega}
\newcommand{\Om}{\Omega}

\newcommand{\fdot}{\mbox{\boldmath $\cdot$}}

\newcommand{\rarrow}{\rightarrow}
\newcommand{\Rarrow}{\Rightarrow}

\newcommand{\nn}{\nonumber}

\begin{document}

\title{Gravito-magnetic instabilities in anisotropically expanding fluids}

\author{Kostas Kleidis$^{1,2}$, Apostolos Kuiroukidis$^{1,3}$, Demetrios B
Papadopoulos$^1$ and Loukas Vlahos$^1$}


\affiliation{$^1$ Department of Physics, Aristotle University of
Thessaloniki, 54124 Thessaloniki, Greece}

\affiliation{$^2$ Department of Civil Engineering, Technological
Education Institute of Serres, 62124 Serres, Greece}

\affiliation {$^3$Department of Informatics, Technological
Education Institute of Serres, 62124 Serres, Greece}

\keywords{Magnetic fields - Cosmological perturbations -
Magnetohydrodynamics - Dynamo effects}

\date{\today}

\begin{abstract}
Gravitational instabilities in a magnetized Friedman - Robertson -
Walker (FRW) Universe, in which the magnetic field was assumed to
be too weak to destroy the isotropy of the model, are known and
have been studied in the past. Accordingly, it became evident that
the external magnetic field disfavors the perturbations' growth,
suppressing the corresponding rate by an amount proportional to
its strength. However, the spatial isotropy of the FRW Universe is
not compatible with the presence of large-scale magnetic fields.
Therefore, in this article we use the general-relativistic (GR)
version of the (linearized) perturbed magnetohydrodynamic
equations with and without resistivity, to discuss a generalized
{\em Jeans criterion} and the potential formation of density
condensations within a class of homogeneous and anisotropically
expanding, self-gravitating, magnetized fluids in curved
space-time. We find that, for a wide variety of anisotropic
cosmological models, {\em gravito-magnetic} instabilities can lead
to sub-horizonal, magnetized condensations. In the non-resistive
case, the {\em power spectrum} of the unstable cosmological
perturbations suggests that most of the power is concentrated on
large scales (small $k$), very close to the horizon. On the other
hand, in a {\em resistive} medium, the critical wave-numbers so
obtained, exhibit a delicate dependence on resistivity, resulting
in the reduction of the corresponding Jeans lengths to smaller
scales (well bellow the horizon) than the non-resistive ones,
while increasing the range of cosmological models which admit such
an instability.
\end{abstract}

\pacs{04.25.Nx, 04.40.Nr, 04.20.Jb}

\maketitle

\section{Introduction}

It is known that, the formation of large-scale structures (such as
galaxies, super-clusters of galaxies and/or super-clouds of stars
within galaxies) both in cosmological and in astrophysical scale
is closely related to the concept of {\em instability}~\cite{1},
\cite {2}.

The first attempt towards a theory of galaxy formation carried out
by Sir James Jeans in the early nineteenths~\cite{3}, \cite{4}. He
assumed that the Universe is filled with a self-gravitating
non-relativistic fluid, characterized by its mass-density $(\rh)$,
the corresponding pressure $(p)$ and a velocity field $(\vec{u})$.
As a consequence, the resulting dynamical system is governed by
the equation of continuity, together with Euler's and Newton's
equations for the matter-content and the gravitational field,
respectively~\cite{5}. However, Jeans' theory does not apply to
the formation of galaxies in an expanding Universe, because it
involves a static medium. Nevertheless, he arrived at the
conclusion that, density fluctuations with wavelengths $\lm$
greater than a {\em critical value} $\lm_J$ will grow, so that the
system, eventually, becomes {\em unstable} (gravitational
instability). As a consequence, any isothermal gaseous sphere with
a length-scale greater than $\lm_J$ is gravitationally unstable
and is going to contract~\cite{6}.

The first satisfactory theory of gravitational instabilities in an
expanding Universe was proposed by Lifshitz~\cite{7}. He showed
that, in a homogeneous and isotropic Universe filled with a
perfect fluid, cosmological perturbations with wave-numbers $(k)$
below the Jeans' one $(k_J)$ grow, not exponentially, but, as a
power of the corresponding scale factor, $R (t)$. Among the
various scenarios of cosmic fluctuations there is also the case
where gravitation and cosmic expansion are essentially irrelevant.
In this context, Weinberg's work~\cite{8} is of particular
interest: For the first time, the notion of an {\em imperfect
fluid} was used and the role of {\em dissipation} in the survival
of protogalaxies was taken into account, along the lines of
Eckart's formalism~\cite{9}, which differs from the corresponding
of Landau \& Lifshitz~\cite{10}. Accordingly, Weinberg found that
the protogalactic fluctuations, which behave as ordinary sound
waves during the period where their matter-content is much lower
than the corresponding Jeans mass $(M << M_J)$, damp in the
acoustic phase due to the {\em viscosity} of the imperfect fluid.

The study of gravitational instabilities in the presence of
magnetic fields in an expanding Universe arose as a natural
generalization~\cite{11}. Accordingly, solutions for the evolution
of the mass-density and the magnetic field $(H^i)$ perturbations
(in the linear regime) have been quested in FRW cosmological
models, in which, the magnetic field was assumed to be too weak to
destroy the isotropy of the model~\cite{12} - \cite{16}.

Although it has been argued that a uniform magnetic field slows
down the growth-rate of the unstable modes, several authors have
studied the influence of cosmological magnetic fields (either
homogeneous or inhomogeneous) to the formation of structures
within a cosmic medium (e.g. see~\cite{17} - \cite{25}, for an
extensive, though incomplete list). In this context, it was
shown~\cite{26} that several magnetohydrodynamic processes like
pinch-effects, hose-instabilities, sausage- and
kink-instabilities, may also play a significant role to structure
formation. These processes correspond to individual harmonics in
the eigenfunction solutions of the wave equation for plasma
instabilities and although they are different, their
time-development is similar.

In the present article, we intend to discuss gravitational
instabilities in a homogeneous and anisotropic cosmological model,
in which the large-scale anisotropy is due to the presence of a
homogeneous magnetic field along the $\hat{z}$-direction.

Our choice is justified by the fact that, mathematically speaking,
the spatial isotropy of the FRW Universe is not compatible with
the presence of large-scale magnetic fields. In fact, an
anisotropic cosmological model can and should be imposed for the
treatment of magnetic fields whose {\em coherent length} is
comparable to the {\em horizon length}~\cite{27}. Therefore,
although current observations give a strong motivation for the
adoption of a FRW model, the effects one may lose by neglecting
the large-scale anisotropy induced by the background magnetic
field, should be investigated. Not to mention that the anisotropy
of the unperturbed model facilitates a closer study of the
coupling between magnetism and geometry.

The Paper is organized as follows: In Section II, we summarize the
general-relativistic version of magnetohydrodynamics (MHD) and
present its application to the Thorne's class of homogeneous and
anisotropic cosmological models~\cite{28}. They correspond to a
class of {\em semi-realistic} solutions to the Einstein-Maxwell
equations, adequate to describe a model which departs from
isotropy along the radiation epoch~\cite{29}. In Section III, we
perturb the corresponding MHD equations to study the evolution of
linear fluctuations propagating along the $\hat{x}$-direction,
i.e. perpendicularly to the magnetic field. An immediate result is
the derivation of the corresponding {\em dispersion relation},
which may reveal the {\em critical values} of the perturbations'
wave-numbers in order to discuss on their potential growth (or
decay). We find that, for a wide class of anisotropic cosmological
models, {\em gravito-magnetic} instabilities can lead to
sub-horizonal, magnetized condensations. In this case, the {\em
power spectrum} of the unstable cosmological perturbations
suggests that most of the power is concentrated on large scales
(small $k$), close to the horizon. These results are in contrast
to what has been found in the isotropic case, where the magnetic
field disfavors large-scale matter aggregation, by reducing the
growth-rate of the corresponding density perturbations \cite{11},
\cite{13}, \cite{14}. In Section IV, we consider the relativistic
MHD equations within a {\em resistive} medium. Now, the critical
wave-numbers so obtained exhibit a delicate dependence on
resistivity, resulting in the reduction of the corresponding Jeans
lengths to smaller scales than the non-resistive ones, while
increasing the range of cosmological models which admit such an
instability.

\section{Magnetohydrodynamics in a model Universe}

In the system of {\em geometrical units}, where both the velocity
of light $(c)$ and Newton's constant $(G)$ are equal to unity, the
GR version of the MHD-equations~\cite{18} reads \bea &&(\rh -
\frac{H^2}{2})_{; \al \bt} \: u^{\al} u^{\bt} = h^{\al \bt}(p +
\frac{H^2}{2})_{; \al \bt} + 2 {d \over dt}(H^2 \theta) \nn \\
&& - (H^{\al} H^{\bt})_{; \al \bt} + 2 \ch \: (\frac{2
\theta^2}{3} + \sg^2 - \om^2 - \dot{u}^{\al} \dot{u}_{\al}) \\
&& + \frac{\ch}{2} (\rh + 3p + H^2) + 2 \dot{u}_{\al}(H^{\al}
H^{\bt})_{; \bt} + (H^2)_{; \al} \dot{u}^{\al} \nn \eea
supplemented by the equation of motion $T^{\al \bt}_{; \: \bt} =
0$ \be \dot{\ch} u^{\al} + \ch \dot{u}^{\al} + \ch \theta u^{\al}
+ (p + \frac{H^2}{2})_{; \bt} \: g^{\al \bt} - (H^{\al}
H^{\bt})_{; \bt} = 0 \ee and Maxwell's equations in curved
space-time \be \dot{H}^{\mu} = (\sg_{\nu}^{\mu} + \om_{\nu}^{\mu}
- \frac{2}{3} \dl_{\nu}^{\mu} \: \theta) H^{\nu} + \frac{1}{\rh +
p} \: p_{, \nu} \: H^{\nu} u^{\mu} \ee In Eqs (1) - (3), the dot
denotes differentiation with respect to $t$, Greek indices refer
to the four-dimensional space-time (in accordance, Latin indices
refer to the three-dimensional spatial section) and
$\dl_{\nu}^{\mu}$ is the Kronecker symbol. Furthermore, $\rh$ is
the mass-energy density as measured in the rest-frame of the
fluid, $p$ is the corresponding pressure and $H^{\mu}$ are the
components of the magnetic field. Accordingly, $\ch$ stands for
the combination $\ch = \rh + p + H^2$, $u^{\al}$ is the fluid's
four-velocity and $h^{\al \bt} = g^{\al \bt} + u^{\al} u^{\bt}$ is
the projection tensor. Finally, $\sg_{\mu \nu}$ and $\om_{\mu
\nu}$ stand for the {\em shear-} and the {\em rotation-} tensor
respectively, while $\theta = u^{\al}_{; \al}$ is the {\em
expansion} parameter, with the semicolon denoting covariant
derivative. Accordingly, $\dot{u}^{\al} = u_{; \bt}^{\al}
u^{\bt}$. The above system is supplemented by the Raychaudhuri
equation~\cite{30} \be \dot{u}_{; \al}^{\al} = \dot{\theta} +
\frac{\theta^2}{3} + 2 (\sg^2 - \om^2) + \frac{1}{2}(\rh + 3p +
H^2) \ee Eqs (1) - (4) govern the evolution of magnetized
plasma-fluids within the context of GR. In what follows, we use
them to describe the propagation of small-amplitude waves
(cosmological perturbations) in a homogeneous and anisotropic
cosmological model, the anisotropy of which is due to the presence
of an ambient magnetic field (frozen into the matter-content)
along the $\hat{z}$-direction, $\vec{H} = H (t) \hat{z}$.

The corresponding line-element is written in the form~\cite{28},
\cite{29}, \cite{31} \be ds^2 = - dt^2 + A^2 (t) (dx^2 + dy^2) +
W^2 (t) dz^2 \ee In Eq (5), the scale factors $A (t)$ and $W (t)$
are solutions to the Einstein-Maxwell equations for a magnetized
perfect-fluid obeying the equation of state \be p = \gm \rh \ee
Therefore, the various models of this class are parameterized by
$\gm$ taking values in the interval $\gm \in [ {1 \over 3} \: , \:
1 )$. Accordingly, $A(t) \sim t^{1/2}$ and $W(t) \sim t^l$, where
$l = (1 - \gm)/(1 + \gm)$~\cite{28}, while \be \rh (\gm, t) =
\frac{3 - \gm}{2 t^2 (1 + \gm)^2} \ee and \be H (\gm ,t) =
\frac{(1 - \gm)^{1/2} (3 \gm - 1)^{1/2}}{2 t (1 + \gm)} \ee

\begin{figure}[h!]
\centerline{\mbox {\epsfxsize=9.cm \epsfysize=7.cm
\epsfbox{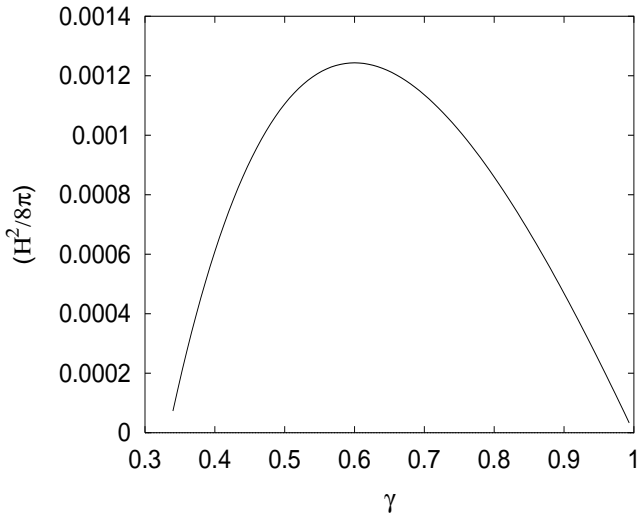}}} \caption{The energy-density
of the magnetic field $(H^2 / 8 \pi)$ in {\it cgs} units, at a
fixed time $(t \simeq 10^{-21} \; sec)$, as a function of $\gm$
taking values in the range $({1 \over 3} \: , \: 1)$. It becomes
maximum in a cosmological model with $\gm = 0.6$.}
\end{figure}

Notice, that for $\gm = {1 \over 3}$ the strength of the magnetic
field vanishes and the model reduces to the radiation-dominated
FRW model. As $t \rarrow 0$, these models evolve towards a
singularity of infinite density and magnetic field. However, for
every ${1 \over 3} < \gm < 1$, the ratio \be {\rh_{magn} \over
\rh_{matt}} = \frac{H^2}{8 \pi \rh} = \frac{(1 - \gm) \: (3 \gm -
1)}{16 \pi \: (3 - \gm)} \ee remains (a non-zero) constant,
suggesting that the influence of the magnetic field to the
evolution of the curved space-time is comparable to that of the
matter.

We assume (e.g. see~\cite{29}) that the metric (5) can describe
the early Universe prior to the {\em recombination epoch}
$(t_{rec} \sim 10^{13} \; sec)$. At $t = t_{rec}$ (i.e. far away
from the initial singularity) the magnetic field is too weak to
have any effect on the dynamics of the model, but preserves the
inherent anisotropy. Therefore, the spatial section of (5) always
remain axially-symmetric. In fact, it is the magnetic field that
accelerates the transverse expansion (over the $xy$-plane), while
decelerating the longitudinal one along the
$\hat{z}$-direction~\cite{28}.

For an observer comoving with the fluid in the space-time (5), we
obtain \be \dot{u}^{\mu} = 0, ~~ \om_{\mu \nu} = 0, ~~ \theta =
\frac{2}{(1 + \gm) t} \ee and \be \sg^2 = \frac{(3 \gm - 1)^2}{6
(1 + \gm)^2 t^2} \ee To study the conditions under which the
background (5) admits gravitational instabilities or not, we turn
to the theory of cosmological perturbations.

\section{Jeans lengths in a magnetized perfect fluid}

For every dynamical system, much can be learnt by investigating
the possible modes of small-amplitude oscillations or waves. A
plasma is physically much more complicated than an ideal gas,
especially when there is an externally applied magnetic field. As
a result, a variety of small-scale perturbations may appear.

Accordingly, we perturb Eqs (1) - (4), introducing (to the
physical variables of the fluid) small-amplitude fluctuations
along the $\hat{x}$-axis and neglect all terms of orders higher or
equal than the second. Within the limits of linear analysis,
excitation of cosmological perturbations in a homogeneous and
anisotropic cosmological model is basically a {\em kinematic
effect}, in the sense that the self-gravitation of the
fluctuations is unimportant (e.g. see~\cite{6}, pp. 501 - 506 and
references therein). In this case, perturbations' growth arises
mostly due to their motion in the anisotropic background.
Therefore, as far as the enhancement of MHD perturbations in the
anisotropic space-time (5) is concerned, we may neglect the
first-order corrections of the metric, admitting the so-called
{\em Cowling approximation}~\cite{32}. In other words, in what
follows, we treat the MHD perturbations as very slowly-varying
(low) frequency waves propagating in an anisotropically expanding
medium, without interacting with the curved space-time, unless the
linear regime breaks down.

To develop the theory of small-amplitude waves in the curved
space-time (5), we follow the so-called {\em adiabatic
approximation}~\cite{1}, \cite{34}, \cite{35} considering that all
the perturbation quantities are proportional to the exponential
\be \exp \: \left [ \imath \left ( k \: x - \int^t n (t^{\prime})
\: d t^{\prime} \right ) \right ] \ee where, $k$ is the comoving
wave-number. In this context, the (slowly-varying) time-dependent
frequency of the wave $[n (t)]$ is defined by the {\em eikonal}
\be \Om = \int^t n (t^{\prime}) \: d t^{\prime} \ee through the
relation $n = d \Om / dt$. Accordingly, we assume a wave-like
expansion of the perturbation quantities, in the form \bea \dl
\rho & = & A_{\rh} e^{\imath (k x - \int^t n d t^{\prime})}, \; \;
\; \; \dl u = A_u e^{\imath (k x - \int^t n d t^{\prime})} \nn \\
\dl H & = & A_H e^{\imath (k x - \int^t n d t^{\prime})} , \; \;
\; \dl p = c_{s}^{2} \: \dl \rh \eea where $c_s^2 = \dl p / \dl
\rh = \gm$ is the {\em speed of sound} (in units of $c$). In the
anisotropic model under consideration, the MHD equations (1) - (4)
linearized over the perturbation quantities, result in
\begin{equation}
\delta u \left [ 2 \ch \: \Gamma_{01}^1 - g^{33} \: \Gamma_{33}^0
\: H^2 + p_{,0}^{*} + \imath n \ch \right ] = \imath k \left [
\gamma \delta \rho + H \: \delta H \right ] \end{equation} and
\begin{equation} \left ( H \: \delta H \right ) \: [(2-\gamma) \:
\theta + \imath n] = \frac{u_A^2}{1 + \gamma} \: [\theta \: (1 +
\gamma)- \imath n] \: \delta \rho
\end{equation} where $\Gm_{\al \bt}^{\gm}$ are the Christoffel
symbols of (5), $g^{\mu \nu}$ are the contravariant components of
the corresponding metric tensor, $u_A^2 = \frac{H^2}{\rh}$ is the
(dimensionless) Alfv\'en velocity [clf Eq (9)] and we have set
$p^{*} = p + \frac{H^2}{2}$.

Before attempting to discuss any temporal evolution of the
perturbation quantities, it is important to trace what kind of
waveforms are admitted by this system. To do so, we have to derive
their {\em dispersion relation}, $D(k, n) = 0$~\cite{35}. Provided
that certain kinds of modes (such as acoustic, magnetosonic etc)
do exist, they can be excited through their interaction with the
anisotropic space-time. An additional excitation, due to the
non-zero resistivity, is also possible~\cite{26}. We have to point
out that, although the background quantities depend on time, in
the search for a dispersion relation, we treat the perturbations'
amplitudes ($A_{i}$s) as {\em constants} (at least initially). In
this way, our search for potential waveforms, is not disturbed by
the inherent non-linearity introduced by the curved background.
Nevertheless, once the potential waveforms are determined, their
interaction with the curved space-time in the presence of an
external magnetic field, implies that, for $t > 0$, the
time-dependence of their amplitudes is {\em a priori} expected
[clf Eq (34), below].

The combination of Eqs (15) and (16) results in the dispersion
relation \be D_r (k, n) + \imath \: D_i (k, n) = 0 \ee the {\em
real part} of which is written in the form \bea D_r (k, n) = & - &
n^2 \: \left [ \frac{22 - 9 \gamma}{9 (2 - \gamma)} \right ] \\ &
+ & k^2 \: \left [ c_s^2 + \frac{u_A^2}{2 - \gamma} \right ] + J_1
- R_3 - \frac{u_A^2}{2 - \gamma} R_1 \nn \eea where, we have set
\begin{equation}\label{dispersionx2}
J_1=-(1+\gamma) \: \left ( \frac{16}{9} \theta^2 + 2 \sigma^2
\right ) - (1 + 3 c_s^2) \: (1+c_s^2) \: \rho
\end{equation} and
\begin{eqnarray}\label{dispersionx3}
&&R_1=-\frac{1}{6}(33\gamma+29) \: \theta^2 + 4 \sigma^2
+ 2 \rho \: (1+2\gamma)+\frac{10}{3}H^2,\nonumber\\
&&R_3=\frac{u_A^2}{9} \left [ (18\gamma-122) \: \theta^2 + 15 \:
(1 + \gamma) \: \rho \right ] \end{eqnarray} The vanishing of Eq
(18) leads to a {\em critical} wave-number, for which the
cosmological perturbations under consideration are {\em stable}
(oscillating); namely, $n^2 \geq 0$ \bea && k^2 \geq k_{r_J}^2
(\gm , t) = \left [- J_1 + R_3 + \frac{u_A^2}{2 - \gamma} R_1
\right ] \left [ c_s^2 + \frac{u_A^2}{2 - \gamma} \right ]^{-1} \nn \\
&& \delta u = 0 \eea As a consequence, for $k^2 < k_{r_J}^2$, the
small-scale fluctuations of the cosmic-medium become {\em
unstable}, acquiring an imaginary frequency $[n_k^2 (t) < 0]$. In
other words, they are no longer oscillating, but evolve
exponentially with time ({\em Jeans-like instability}).

In the limiting case where $\gm = {1 \over 3}$ (FRW radiation
model), we have $H = 0 = u_A^2$, $\sg^2 = 0$, $\theta \neq 0$ and
Eq (18) takes the simpler form obtained in~\cite{16}. In this
case, a Jeans instability arises at wave-numbers $k^2 < k_{r_J}^2
({1 \over 3} , t)$, where
\begin{equation}\label{dispersionf1}
k_{r_J}^2 ({1 \over 3} , t) \: c_s^2 = (1 + c_s^2) \: (1 + 3
c_s^2) \: \rho + \frac{16}{9} (1 + c_s^2) \: \theta^2
\end{equation} On the other hand, for $\gm \rarrow 1$ (Zel'dovich
ultra-stiff matter) we obtain $H = 0 = u_A^2$, while $\sg^2 \neq
0$, $\theta \neq 0$ and Eq (18) also yields the corresponding
result of~\cite{16}; namely,
\begin{equation}\label{dispersionf3}
k_{r_J}^2 (1 , t) \: c_s^2 = (1 + c_s^2) \: (1 + 3 c_s^2) \: \rho
+ \frac{4}{3} \: \left ( \frac{16}{9} \: \theta^2 + 2 \sigma^2
\right ) \end{equation} Finally, in the extreme case of a static
model, i.e. $\theta = 0$ and $\sg = 0$, we recover the well-known
Jackson's solution~\cite{36}. Accordingly, we conclude that, the
critical wave-number (21) possesses the correct asymptotic
behavior in every limiting case.

The rhs of Eq (21) is a function of time, as well as of $\gm$. In
particular, \be k_{r_J}^2 (\gm , t) = {P_5 (\gm)/ Q_3 (\gm) \over
(\gm + 1)^2 \: t^2} \ee where, for all $\gm$ in the range $[{1
\over3} \: , \: 1)$, \bea Q_3 (\gm) & = & 36 \: \gm \: (3 - \gm) \: (2 - \gm) \nn \\
& + & 18 \: (1 - \gm) \: (3 \gm - 1) > 0 \eea and \bea P_5 (\gm) &
= & 2 (1 + \gm) (3 - \gm) (2 - \gm) (27 \gm^2 +36 \gm + 161) \nn
\\ & + & (1 - \gm) (3 \gm -1) (2 - \gm) (-15 \gm^2 + 174 \gm -
931) \nn \\ & + & 3 (1 - \gm) (3 \gm -1) (9 \gm^2 -106 \gm -99) >
0 \eea i.e. they are (positive) polynomials of $\gm$, of the
third- and the fifth-order, respectively (see Fig. 2). We observe
that, for every ${1 \over 3} \leq \gm < 1$, we have \be k_{r_J}^2
(\gm , t) \leq k_{r_J}^2 ({1 \over3} , t) \ee and therefore  \be
\lm_{r_J} (\gm , t) \geq \lm_{r_J} ({1 \over 3} , t) \ee In other
words, the more we depart from a FRW radiation model, the larger
is the scale at which gravito-magnetic instabilities can become
prominent. Departure from $\gm = {1 \over 3}$, suggests that the
initial fortification of the magnetic-field strength - for ${1
\over 3} \leq \gm \leq 0.6$ - is compensated by a subsequent
decrease - for $0.6 < \gm < 1$ (in connection see Fig. 1).
However, in any case, the corresponding length-scale grows - at
least up to $\gm \lesssim 0.85$, after which it remains
practically constant (Fig. 2). Therefore, we may conclude that,
any non-vanishing ambient magnetic field disfavors small-scale
(sub-horizonal) condensations within a non-resistive plasma-fluid.
This result is compatible to what has already been found in the
isotropic case (e.g. see~\cite{13}, \cite{14}).

\begin{figure}[h!]
\centerline{\mbox {\epsfxsize=9.cm \epsfysize=7.cm
\epsfbox{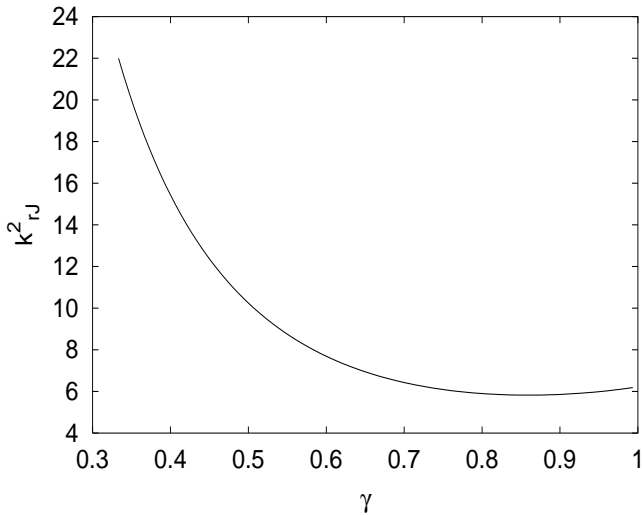}}} \caption{The critical
wave-number $k_{r_J}^2$ (in {\it cgs} units) at a fixed time $(t
\simeq 10^{-21} \; sec)$, as a function of $\gm$ taking values in
the range $[{1 \over 3} \: , \: 1)$.}
\end{figure}

Since $k_{r_J} \sim t^{-1}$, the entirety of critical wave-numbers
in the form (24) vanish only at $t \rarrow \infty$. Hence, in the
curved space-time (5) we may always find a length-scale involving
a {\em gravito-magnetic instability}. However, the MHD
perturbations considered, correspond to low-frequency waves and,
hence, these results should be valid (mainly) at the late stages
of the radiation epoch (probably around {\em recombination}). For
every $t$ within the radiation era, we may use Eqs (24) - (26) to
estimate the corresponding critical length-scales (see also Fig.
2). In accordance, \be k_{r_J} \: \simeq \: (2.5 \: - \: 4.7)
\times {1 \over t} \ee (in geometrical units) where, the higher
values correspond to $\gm$ in the vicinity of ${1 \over 3}$ (in
particular, $\gm \lesssim 0.5$), while, the lower ones to the rest
of the range. The corresponding Jeans length reads \be \lm_{r_J} =
{2 \pi \over k_{r_J}} \: \simeq \: (1.3 \: - \: 2.5) \times t \ee
in geometrical units or \be \lm_{r_J} \: \simeq \: (4 \: - \: 7.5)
\times 10^{10} \times t \ee in {\it cgs} units. It is easy to
verify that, for all $t$ within the radiation era, these scales
are comparable to the horizon size along the $xy$-plane ($\ell_{H
\perp} = 2 c t = 6 \times 10^{10} \times t$, in {\it cgs} units).
In fact, \be 0.67 \: \ell_{H \perp} \leq \lm_{r_J} \leq 1.25 \:
\ell_{H \perp} \ee with the lower values (i.e. those for which
$\lm_{r_J} \lesssim \ell_{H \perp}$) corresponding to $\gm
\lesssim 0.5$ and the higher ones to $0.5 < \gm < 1$. Confining
ourselves to sub-horizonal scales (e.g. see~\cite{37}), we may
conclude that, as regards {\em favorism} of gravito-magnetic
instabilities in a cosmological model of the form (5), the values
of $\gm > 0.5$ should be discarded.

As regards the sub-horizonal perturbations, for a fixed $k$, Eqs
(21) and/or (24) introduce to the evolution of the corresponding
mode a {\em critical} time-scale $(t_{k_{r_J}})$ in the following
sense: As long as \be t \: < \: {\sqrt {P_5 (\gm) / Q_3 (\gm)}
\over (\gm + 1) \: k} = t_{k_{r_J}} \ee the cosmological
perturbation involved is unstable, i.e. $n_k^2 (t < t_{k_{r_J}}) <
0$ and grows exponentially with time~\cite{38} \bea &&\dl \rh (t)
\: , \: \: \dl H (t) = A_i \: \exp \left ( - \imath \int^t n_k
(t^{\prime}) \: d t^{\prime} \right ) \nn \\
& = & A_i \: \left [ {\sqrt {P_5} - \sqrt {P_5 \: - \: (\gm +1)^2
Q_3 \: k^2 \: t^2} \over \sqrt {P_5} + \sqrt {P_5 \: - \: (\gm
+1)^2 Q_3 \: k^2 \: t^2}} \right ]^{- {1 \over 2} a (\gm) \sqrt
{P_5}} \nn \\ & \times & \exp \left [ - a (\gm) \sqrt {P_5 \: - \:
(\gm +1)^2 Q_3 \: k^2 \: t^2} \right ] \eea where $A_i$ stands
either for $A_{\rh}$ or $A_H$, \be a (\gm) = {1 \over 2 (\gm +1)
\sqrt{(22 - 9 \gm) (3 - \gm)}} \ee and the exponentially decaying
solutions - those with the {\it plus} sign in the exponents - are
neglected. With the aid of Eq (24), Eq (34) is written in the form
\bea \dl \rh (k) \: , \: \: \dl H (k) & = & A_i \: \left [ {1 -
\sqrt {1 - ({k \over k_{r_J}})^2} \over 1 + \sqrt {1 - ({k \over
k_{r_J}})^2}} \right ]^{- {1 \over 2} a (\gm) \sqrt {P_5 (\gm)}}
\nn \\ & \times & e^{- a (\gm) \sqrt {P_5 (\gm)} \: \sqrt {1 - ({k
\over k_{r_J}})^2}} \eea In the limit $k \rarrow k_{r_J}$, Eq (36)
yields $\dl \rh \: , \: \dl H = A_i$ (the constant amplitude of an
oscillating perturbation), while, for $k \ll k_{r_J}$ we obtain
\be \dl \rh (k) \: , \: \: \dl H (k) \gtrsim A_i \: \left
({k_{r_J} \over k} \right )^{a (\gm) \sqrt {P_5 (\gm)}} > A_i \ee
Accordingly, condensations of the corresponding length-scale
$(\lm_{r_J})$ will collapse in Jeans' fashion, while, at the same
time, the fluctuations of the magnetic field will grow, resulting
in a {\em gravito-magnetic instability}, as well as in the
amplification of the background magnetic field (in connection,
see~\cite{25}, \cite{39}).

The behavior of the (large scale) magnetized perturbations in the
anisotropic model under consideration, is completely different
than their behavior in the isotropic FRW model. In that case, both
the Newtonian~\cite{11} and the GR study \cite{13}, \cite{14} have
verified a negative role of the magnetic field in the aggregation
of matter on large scales. In particular, it has been shown that,
on scales between the Jeans length and the horizon, the density
perturbations may undergo a power-law evolution similar to that of
the non-magnetized case (e.g. see \cite{1}), but their growth-rate
is reduced by an amount proportional to the magnetic field's
strength. Therefore, in the isotropic case, the external field
disfavors viable matter condensations.

As regards the anisotropic case, the density perturbation at any
spatial location can be obtained by a superposition of modes with
different wave-numbers \be \dl \rh (\vec{x} \: , \: t) = \int {d^3
k \over (2 \pi)^3} \: \dl \rh (k) \: e^{\imath \vec{k} \fdot
\vec{x}} \ee The strength of such a perturbation can be measured
by the value of $\vert \dl \rh (\vec{x} , t) \vert^2$. From Eq
(38), it follows that modes in the range $(k \: , \: k + dk)$
contribute to $\vert \dl \rh (\vec{x} , t) \vert^2$ an amount
proportional to $\vert \dl \rh (k) \vert^2 d^3 k$. Writing the
contribution as $$\vert \dl \rh (k) \vert^2 \: d^3 k \: = 4 \pi \:
\vert \dl \rh (k) \vert^2 \: k^2 \: dk \: = 4 \pi \: k^3 \: \vert
\dl \rh (k) \vert^2 \: d (\ln k)$$ we see that each logarithmic
$k$-interval contributes to $\vert \dl \rh (\vec{x} , t) \vert^2$
an amount ${\cal P}_k^2 \sim k^3 \vert \dl \rh (k) \vert^2$, which
represents the so-called {\em power spectrum}~\cite{1}. With the
aid of Eq (37), the power spectrum of the MHD perturbations under
consideration is written in the form \be {\cal P}_k^2 \sim k^{3 -
2 a (\gm) \sqrt {P_5 (\gm)}} \ee which, for $\gm = {1 \over 3}$
reads ${\cal P}_k \sim k^{-0.9}$, while, in the limiting case $\gm
\rarrow 1$, it results in ${\cal P}_k \sim k^{-0.6}$. Notice that,
although both cases correspond to the absence of the ambient
magnetic field, the {\em spectral indices} involved are different.
This difference arises due to the {\em stiffness} of the Universe
matter-content as $\gm \rarrow 1$ ($c_s \rarrow c$). The explicit
form of the spectral index \be \nu (\gm) = {k \over \dl \rh} \: {d
\over d k} \dl \rh = {3 \over 2} - a (\gm ) \sqrt {P_5 (\gm)} \ee
as a function of $\gm \in [{1 \over 3} \: , \: 1)$, is presented
in Fig. 3. We note that, for every $\gm$ in this range, we have
$\nu (\gm) < 0$. Once again, we verify that most of the {\em
power} is concentrated on large scales (small wave-numbers).

\begin{figure}[h!]
\centerline{\mbox {\epsfxsize=9.cm \epsfysize=7.cm
\epsfbox{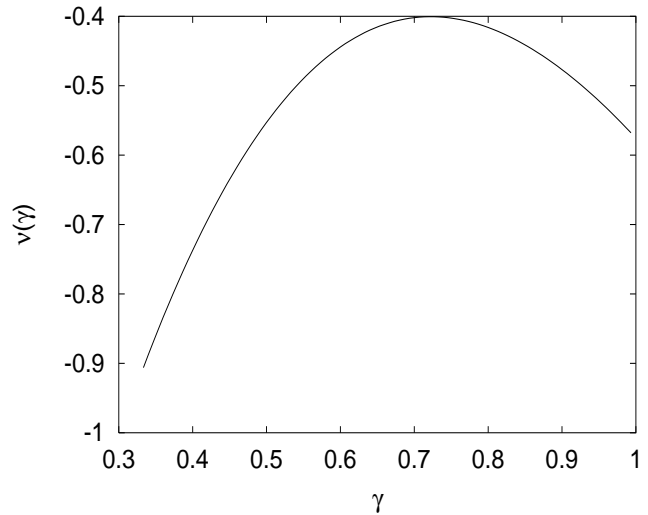}}} \caption{The spectral index
$\nu (\gm)$ of the unstable perturbations as a function of $\gm$,
taking values in the range $[{1 \over 3} \: , \: 1)$.}
\end{figure}

\section{Jeans lengths in a resistive plasma}

It has been argued~\cite{26} that, prior to recombination, the
{\em conductivity} of the Universal plasma fluid is finite,
although it may reach at very high values, of the order $10^{12} -
10^{13} \; m/ohm)$. Accordingly, the study of MHD perturbations in
a Bianchi Type I model has indicated that the finite conductivity
(non-zero {\em resistivity}) could result in an exponential growth
of the density perturbations. This argument was accordingly
confirmed in the Newtonian limit~\cite{11}, but not in the
relativistic one. In this Section, we shall attempt to re-examine
the above statements in the relativistic limit. The equations
which govern the hydrodynamics of a magnetized, resistive fluid in
the frame of GR, are identical to Eqs (1) and (2) being
supplemented by the Maxwell equations
\begin{equation}\label{cmhd3}
\dot{H}^{\alpha}=(\sigma_{\beta}^{\alpha}+\omega_{\beta}^{\alpha})
H^{\beta}-\frac{2}{3}\theta
H^{\alpha}+u^{\alpha}H_{;\gamma}^{\gamma} + \et \: h^{\gamma
\beta}H_{;\gamma\beta}^{\alpha}
\end{equation}
which coincide to Eqs (3), with the addition of an extra term due
to the resistivity $(\et)$ of the cosmic medium.

In this case, as regards the propagation of cosmological
perturbations along the $\hat{x}$-direction, Eq (16) reads
\begin{equation} (H \; \delta H) \: [(2 - \gamma) \: \theta + k^2 \eta +
\imath n] = \frac{u_A^2}{1+\gamma} \: [\theta \: (1+\gamma) -
\imath n] \: \delta \rho \end{equation} Accordingly, the vanishing
of the {\em real part} of the dispersion relation results in \bea
&&n^2 \left [ {22 - 9 \gm \over 9} \: \theta + k^2 \et \right ] =
(2 - \gm) \left \{ k^4 \: {c_s^2 \et \over 2 - \gm} \right . \nn
\\ & + & k^2 \: \left [ (c_s^2 + {u_A^2 \over 2 - \gm}) \: \theta
+ {\et \over 2 - \gm} (J_1 - R_3) \right ] \nn \\ & + & \left .
(J_1 - R_3 - {u_A^2 \over 2 - \gm} \: R_1) \: \theta \right \},
~~~ \delta u = 0 \eea Eq (43) can be cast in the form
\begin{equation}\label{cdispersionx1} n^2 \: \left [ {22 - 9 \gm
\over 9} \: \theta + k^2 \et \right ] = k^4 m_4+k^2 m_2+m_0
\end{equation} where, for every $\gm \in [{1 \over 3} \: , \: 1)$,
\bea m_4 & = & c_s^2 \: \eta \: > \: 0 \nn \\
m_0 & = & (2-\gamma) \: \theta \: (J_1-R_3)-\theta \: u_A^2 \: R_1
\nn \\ & = & - \: {P_5 (\gm) \over 18 \: (3 - \gm) \: (1 + \gm)^3
\: t^3} \: < \: 0 \eea and \bea m_2 & = & c_s^2 \: (2
- \gamma) \: \theta + \theta \: u_A^2 + (J_1 - R_3) \: \eta \\
& = & {Q_3 (\gm) \over 18 (3 - \gm) (1 + \gm) \: t} - \et \: {B_4
(\gm) \over 36 (3 - \gm) (1 + \gm)^2 \: t^2} \nn \eea while \bea
B_4 (\gm) & = & 2 (1 + \gm) (3 - \gm) \left (27 \gm^2 + 36 \gm +
161 \right ) \\ & + & (1 - \gm) (3 \gm - 1) \left ( -15 \gm^2 +174
\gm - 931 \right ) > 0 \nn \eea

The allowed wave-numbers for stable (oscillating) perturbations
can be easily obtained by Eq (44), admitting that $n^2 \geq 0$.
Accordingly, \begin{equation} \label{ckxr} m_4k^4+m_2k^2+m_0 \geq
0 \end{equation} Since $m_2^2 - 4 m_4 m_0 > 0$, the trinomial on
the lhs of Eq (48) possesses the couple of real roots
\begin{equation}\label{ckxr12} k_{r_{J_{(1,2)}}}^2 = \frac{-m_2}{2 m_4}
\pm \frac{1}{2m_4} \sqrt{m_2^2-4m_4 m_0} \end{equation} For small
values of $\et$ $(\et \rarrow 0)$, i.e. within the context of an
{\em ideal}-plasma cosmology, the corresponding wave-numbers
behave as \bea k_{r_{J_{(1,2)}}}^2 \simeq & - & {1 \over 36 \: \et
\: \gm} \: {Q_3 (\gm) \over (3 - \gm) (1 + \gm) \: t} \nn
\\ & \pm & {1 \over 36 \: \et \: \gm} \: {Q_3 (\gm) \over (3 - \gm)
(1 + \gm) \: t} \nn \\ & \pm & {P_5 (\gm) / Q_3 (\gm) \over (1 +
\gm)^2 \: t^2} \eea and therefore, now, the condition for {\em
unstable} perturbations reads \bea k_{r_{J_1}}^2
\leq k^2 & \leq &  k_{r_{J_2}}^2 \Rarrow \nn \\
0 < k^2 & \leq & k_{r_{J_2}}^2 = {P_5 (\gm) / Q_3 (\gm) \over (1 +
\gm)^2 \: t^2} \eea i.e. it is identical to the corresponding
non-resistive one.

On the other hand, in the limit of large $\et$, we obtain \bea
k_{r_{J_{(1,2)}}}^2 & = & {B_4 (\gm) \over 72 \: \gm \: (3 - \gm)
(1 + \gm)^2 \: t^2} \\
& \pm & {B_4 (\gm) \over 72 \: \gm \: (3 - \gm) (1 + \gm)^2 \:
t^2} \; + \; O \: ({1 \over \et}) \nn \eea In accordance, as long
as \be 0 < k^2 \leq k_{r_{J_2}}^2 = {B_4 (\gm) \over 36 \: \gm \:
(3 - \gm) (1 + \gm)^2 \: t^2} \ee the condition (48) {\it is not}
fulfilled and the corresponding cosmological fluctuations are
unstable and grow exponentially with time $(n_k^2 < 0)$. As a
consequence, condensations of the corresponding length-scale will
collapse in the Jeans' fashion, revealing an indirect {\em
resistive instability} [the rhs of Eq (53) is independent of
$\et$].

In what follows, we compare the length-scales arising by the
critical wave-numbers of the resistive and the non-resistive case.
Our aim is to examine the contribution of resistivity in achieving
more realistic (sub-horizonal) scales of gravito-magnetic
instability, than those predicted by the non-resistive fluid. To
do so, we consider the ratio of the critical wave-numbers (53) and
(24), which is written in the form \be {k_{res}^2 \over k_{non}^2}
= {B_4 (\gm) \: Q_3 (\gm) \over 36 \: \gm \: (3 - \gm) \: P_5
(\gm)} \ee and is depicted in Fig. 4. First of all, we observe
that for $\gm = {1 \over 3}$ and/or $\gm = 1$, i.e. in the absence
of the ambient magnetic field, we obtain $k_{res} = k_{non}$. This
not an unexpected result, since $\et$ is introduced to the MHD
analysis through the Maxwell equation (41), which involves a
non-vanishing magnetic-field strength. On the other hand, a large
resistivity can modify quite significantly the scales involved in
a gravito-magnetic instability and especially those around the
value $\gm = {2 \over 3}$ (the peak in Fig. 4), where the
non-resistive analysis predicts a super-horizonal Jeans length
$[\lm_{non} (\gm = {2 \over 3} , t) = 1.20 \: \ell_{H \perp}
(t)]$. Accordingly, for $\gm = {2 \over 3}$, we obtain \be k_{res}
\simeq 1.2 \: k_{non} \Rarrow \lm_{res} \simeq 0.83 \: \lm_{non}
\lesssim \ell_{H \perp} (t) \ee Therefore, we may conclude that,
as compared to the ideal-plasma case, gravito-magnetic
instabilities within a magnetized-fluid are particularly favored
by a large resistivity, in the sense that it prolongs the range of
the potential values of $\gm$ (and therefore the number of the
corresponding cosmological models) which admit sub-horizonal
scales of such an instability (from ${1 \over 3} \leq \gm \lesssim
{1 \over 2}$ in the non-resistive case, to ${1 \over 3} \leq \gm
\lesssim {2 \over 3}$ in the fully-resistive one).

\begin{figure}[h!]
\centerline{\mbox {\epsfxsize=9.cm \epsfysize=7.cm
\epsfbox{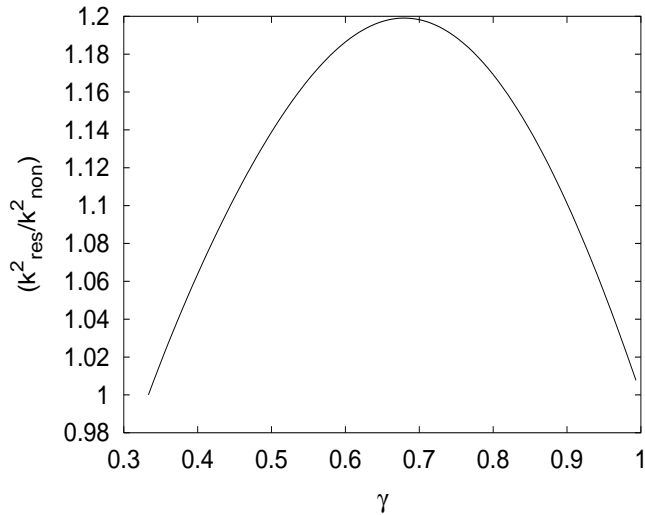}}} \caption{The ratio
$k_{res}^2 / k_{non}^2$ of the unstable perturbations as a
function of $\gm$, taking values in the range $[{1 \over 3} \: ,
\: 1)$.}
\end{figure}

This result, which is based on the fact that the critical (Jeans)
wave-number in the resistive case is always larger than the
corresponding non-resistive one (clf Fig. 4), has a clear physical
interpretation: Indeed, the invariant form of Ohm's law for a
locally neutral plasma is written as \cite{35} \be J^{\mu} = {1
\over \et} \: F^{\mu \nu} u_{\nu} \ee where, $J^{\mu}$ is the
current density and $F^{\mu \nu}$ is the Faraday tensor. According
to it, besides the {\em convective field}, a non-zero resistivity
favors also {\em convective currents}. As a consequence, for $\et
\neq 0$, the energy available to be absorbed by the perturbation
quantities is larger than that of the non-resistive case. This
energy-excess will lead to the increase of the perturbations'
frequency and hence to the amplification of their wave-number, as
well (see also \cite{25}).

\section{Conclusions}

Gravitational instabilities in a magnetized FRW Universe, in which
the magnetic field was assumed to be too weak to destroy the
isotropy of the model, are known and have been studied in the
past~\cite{13}, \cite{14}, \cite{16}. However, mathematically
speaking, the spatial isotropy of the FRW Universe is not
compatible with the presence of large-scale magnetic fields. In
fact, an anisotropic cosmological model can and should be imposed
for the treatment of magnetic fields whose coherent length is
comparable to the horizon length~\cite{27}. Interesting
homogeneous and anisotropic cosmological models of Bianchi Type I
with an ambient magnetic field are known~\cite{28} and have been
used to discuss astrophysical implications~\cite{6}, \cite{19},
\cite{25}, \cite{29}, \cite{31}.

In this article, we have used the GR version of the MHD
equations~\cite{18} in a homogeneous and anisotropically
expanding, self-gravitating, magnetized plasma, to study the
gravito-magnetic excitation of cosmological perturbations.
Accordingly, we have found that:

\begin{itemize}

\item In a non-resistive medium:

\begin{enumerate}

\item In contrast to the isotropic case, where the role of the
magnetic field (as an agent opposing the perturbations' growth) is
clear, there is a wide class of anisotropic, magnetized
cosmological models - parameterized by $\gm$ in the range ${1
\over 3} \leq \gm \lesssim 0.5$, for which a {\em
gravito-magnetic} instability can lead to sub-horizonal magnetized
condensations [clf Eqs (29), (31) and (32)].

\item We have determined explicitly the {\em power spectrum} of
the unstable cosmological perturbations [Eq (37)], which indicates
that most of the power is concentrated on large scales, close to
the horizon [clf Eq(40)].

\end{enumerate}

\item In a resistive medium:

\begin{enumerate}

\item The critical wave-numbers so obtained, exhibit a delicate
dependence on resistivity [Eq (52)], resulting in the reduction of
the corresponding Jeans lengths to smaller scales (well below the
horizon) than the non-resistive ones. In fact, the non-zero
resistivity induces convective currents, resulting in the
enhancement of the perturbations' energy (frequency) and hence in
the reduction of their comoving wavelength.

\item Furthermore, the non-zero resistivity prolongs the range of
the potential values of $\gm$ (and therefore the number of the
corresponding cosmological models) which admit sub-horizonal
scales of such an instability (from ${1 \over 3} \leq \gm \lesssim
{1 \over 2}$ in the non-resistive case, to ${1 \over 3} \leq \gm
\lesssim {2 \over 3}$ in the fully-resistive one) [clf Eq (54)].

   \end{enumerate}

\end{itemize}

\begin{acknowledgements}
The authors would like to thank Dr Heinz Ishliker and Dr Christos
G. Tsagas for several helpful discussions. We also thank the
anonymous referee for his critical comments and his useful
suggestions, which greatly improved the article's final form.
Finally, financial support from the Greek Ministry of Education
under the Pythagoras programm, is also gratefully acknowledged.
\end{acknowledgements}

\section*{Appendix A}

The vanishing of the {\em imaginary part}, $D_i (k , n)$, of the
dispersion relation (17) results in $$n^2 \left [ 1 +
\frac{u_A^2}{1 + \gamma} \right ] =  k^2 \left [ c_s^2 -
\frac{u_A^2}{1 + \gamma} \right ] + \frac{4}{9} (\theta - R_4) (2
- \gamma) \: \theta +$$\\$$+ J_1 - R_3 + \frac{u_A^2}{1 + \gamma}
\left [ R_1 + \theta R_2(1+\gamma) \right ] \eqno{(A1)}$$ where
$$R_2 = \frac{1}{2} \: \theta \: (3\gamma+1) \eqno{(A2)}$$ and
$$R_4 = - \frac{u_A^2}{2(1+\gamma)} \: \theta \: (18\gamma-122)
\eqno{(A3)}$$ In this case, the requirement $n^2 \geq 0$ for
oscillating ({\em stable}) fluctuations leads to the {\em
critical} (Jeans) wave-number $$k_{i_J}^2 \left [ c_s^2 - {u_A^2
\over 1 + \gm} \right ] = - \frac{4}{9} (\theta - R_4) (2 -
\gamma) \: \theta -$$\\$$- J_1 + R_3 - \frac{u_A^2}{1+\gamma}
\left [ R_1 + (1 + \gamma) \: \theta \: R_2 \right ] \eqno{(A4)}$$
Once again, for every $\gm \in [{1 \over 3} \: , \: 1)$ the rhs of
Eq (A4) is of the form (24), thus remaining positive for all $t$.
Therefore, in a non-resistive fluid, the small-scale cosmological
perturbations with $k < k_{i_J}$ are unstable and grow
exponentially with time. Hence, formations of the corresponding
length-scale will collapse as a result of a gravito-magnetic
instability.

On the other hand, as regards the propagation of MHD fluctuations
in a resistive fluid, the vanishing of the {\em imaginary part} of
the corresponding dispersion relation yields $$n^2 \left [ 1 +
\frac{u_A^2}{1 + \gamma} \right ] = k^2 \left [c_s^2 -
\frac{u_A^2}{1 + \gamma} + \eta \: \frac{4}{9} (\theta - R_4 )
\right ] +$$\\$$+ J_1 - R_3 + \frac{4}{9} (\theta - R_4) (2 -
\gamma) \: \theta +$$\\$$+ \frac{u_A^2}{1+\gamma}[R_1 + R_2 \:
\theta \: (1+\gamma)] \eqno{(A5)}$$ Accordingly, for $n^2 \geq 0$
we obtain $k^2 \geq k_{i_J}^2$, where $$ k_{i_J}^2 = \left [ c_s^2
- \frac{u_A^2}{1 + \gamma} + \eta \: \frac{4}{9} (\theta - R_4 )
\right ]^{-1} \times$$\\$$\times \left \{-J_1 + R_3 - \frac{4}{9}
(\theta - R_4) \: (2-\gamma) \: \theta \right . -$$\\$$- \left .
\frac{u_A^2}{1+\gamma} \: [R_1+R_2 \: \theta \: (1 + \gamma) ]
\right \} \eqno{(A6)}$$ In this case, the critical wave-numbers so
obtained $(k_{i_J}^2)$, are weakly-dependent on resistivity. In
particular, around recombination the resistivity is related to the
background temperature $(T)$ according to~\cite{11}, \cite{40},
\cite{41} $$\et = \et_n \: T^{-3/2} \eqno{(A7)}$$ where $\et$ (in
geometrical units) is measured in {\it sec}. At the late stages of
the radiation epoch $T = 2.7 \: (1+z)$, where $z$ is the
cosmological redshift, while $\et_n = 7 \times 10^{10}$ for a
weakly ionized gas and $\et_n = 10^{13}$ for the fully ionized
one. Therefore, at $z_{rec} \simeq 1500$, the product $\et \times
\theta$ (in geometrical units) is dimensionless, admitting the
value $\et \times \theta \sim 2 \times \frac{10^{-8}}{( 1 + \gm)}$
(for a weakly ionized gas). For this reason, the influence of
resistivity in Eq (A6), as compared to Eq (A4), is of minor
interest, since it simply adds a term of the order $10^{-8}$ to
the evaluation of $k_{i_J}^2$.

\end{document}